\documentclass[fleqn,usenatbib]{mnras}

\usepackage{newtxtext,newtxmath}

\usepackage[T1]{fontenc}

\DeclareRobustCommand{\VAN}[3]{#2}
\let\VANthebibliography\thebibliography
\def\thebibliography{\DeclareRobustCommand{\VAN}[3]{##3}\VANthebibliography}

\usepackage{graphicx}	
\usepackage{amsmath}	
\usepackage{multirow}   
\usepackage[dvipsnames]{xcolor}

\title[Large grains and a dust trap in HD~98800B]{Large Dust Grains and a Possible Dust Trap in the Polar Circumbinary Disc of HD~98800B}

\author[Á. Ribas et al.]{
Álvaro Ribas,$^{1,2}$\thanks{E-mail: a.ribas@qmul.ac.uk}
Thomas Lack,$^{2}$
Francesco Zagaria,$^{3}$
Enrique Macías,$^{4}$
Sean M. Andrews,$^{5}$
Amelia Bayo,$^{4}$
\newauthor
Cathie J. Clarke,$^{2}$
Nicolás Cuello,$^{6}$
and Catherine C. Espaillat$^{7,8}$
\\
$^{1}$Astronomy Unit, Department of Physics and Astronomy, Queen Mary University of London, Mile End Road, London E1 4NS, UK\\
$^{2}$Institute of Astronomy, University of Cambridge, Madingley Road, Cambridge, CB3 0HA, UK\\
$^{3}$Max Planck Institute for Astronomy, Königstuhl 17, 69117 Heidelberg, Germany\\
$^{4}$European Southern Observatory, Karl-Schwarzschild-Stra\ss e 2, 85748 Garching bei M\"{u}nchen, Germany\\
$^{5}$Center for Astrophysics \textbar\ Harvard \& Smithsonian, 60 Garden Street, Cambridge, MA 02138, USA\\
$^{6}$University Grenoble Alpes, CNRS, IPAG, 38000 Grenoble, France\\
$^{7}$Department of Astronomy, Boston University, 725 Commonwealth Avenue, Boston, MA 02215, USA\\
$^{8}$Institute for Astrophysical Research, Boston University, 725 Commonwealth Avenue, Boston, MA 02215, USA
}

\date{Accepted 2026 March 24. Received 2026 March 24; in original form 2025 December 11}

\pubyear{2025}

\begin{document}
\label{firstpage}
\pagerange{\pageref{firstpage}--\pageref{lastpage}}
\maketitle

\begin{abstract}
     HD~98800 is a nearby hierarchical quadruple system comprising two binaries orbiting each other. Surprisingly, despite its $\sim 10$~Myr age and dynamic environment, the Ba-Bb component is surrounded by a compact gas-rich disc in a polar configuration. Previous millimetre continuum observations of this disc found a low millimetre spectral index ($\alpha \sim 2.1$ up to 9~mm), potentially arising from large dust grains, optically thick emission, or both. Furthermore, the interpretation was complicated by emission mechanisms other than dust thermal continuum at longer wavelengths. We present new observations of this system with the Very Large Array (VLA) at 6.8~mm and 3~cm, providing crucial additional sampling of the emission at millimetre/centimetre wavelengths. By combining these with ancillary data, we derive a dust spectral index $\alpha_{\rm dust} < 3$ for wavelengths $\le 1$~cm. Our modeling suggests that the emission is optically thick at short millimetre wavelengths ($\lambda \le 3$~mm) and it becomes at least partially optically thin for the VLA observations. The shallow spectral index thus indicates the existence of large grains in the disc. We also identify gyro-synchrotron emission from the A and B components at $\lambda \gtrsim $3~cm. The VLA images also reveal an azimuthal asymmetry at 6.8~mm and 8.8~mm, which is not present in high-resolution ALMA 1.3~mm data. After ruling out geometric and illumination effects, we interpret this asymmetry as a local dust overdensity, possibly induced by a vortex or a relic of the previous passage of the A component.
\end{abstract}




\section{Introduction}

Planet formation is a complex process that involves the growth of dust particles from their original sub-$\mu$m sizes in the interstellar medium, to millimetre/centimetre-sized pebbles, and eventually into planetesimals \citep{Drazkowska2023}. This process takes place in protoplanetary discs which are ubiquitous around young stars. However, in multiple stellar systems, planet formation faces additional complications such as a changing gravitational field, disc truncation, or increased collisional velocities between dust particles and planetesimals \citep{Jang-Condell2015,Pierens2021,Cuello2025,Alaguero2025}. Evidence suggests that, in fact, binary systems present a lower planet occurrence rate than single stars \citep[e.g.,][]{Lester2021,Sullivan2026}, possibly a result of these challenges. On the other hand, planets {\it do} exist in binaries and multiple stellar systems \citep{Doyle2011,Martin2018,Baycroft2025,Zuniga2025,Thebault2025}. Given that a significant fraction of stars in the Galaxy start their lives in such systems \citep{Offner2023}, a complete understanding of planet formation must include them too.

HD~98800 offers a privileged laboratory for these studies. Located at just 45~pc and with an estimated age of $\sim$10~Myr \citep[it is a member of the TW~Hya association,][]{Soderblom1998,Torres2008}, it is a hierarchical quadruple system consisting of two binary systems (A and B), orbiting each other with separation $a=50-70$~au \citep[][]{Boden2005,Kennedy2019,Zuniga2021}. Each of those binary systems contains a binary pair (Aa-Ab and Ba-Bb), both on close and eccentric ($\sim$1~au, $e=0.5-0.8$) orbits \citep[][]{Boden2005,Zuniga2021}. In addition, the B (Ba-Bb) component is surrounded by a compact gas-rich circumbinary disc extending from 2.5 to 4.6~au in dust continuum emission \citep{Andrews2010a,Ribas2018,Kennedy2019}, while the A (Aa-Ab) component appears to be discless. Recent scattered light observations taken with SPHERE also reveal a very compact disc \citep[$\sim$3~au radius,][]{Engler2025}.\footnote{The SPHERE data show two dips in scattered light at ${\rm PA}=109\deg$ and ${\rm PA}=280\deg$. Although it is tempting to interpret them a shadow, they are possibly due to the complex removal of the binary signal \citep{Engler2025}.} The existence of the circumbinary disc in HD~98800B is surprising given the system's age and dynamic environment: \citet{Ribas2018} proposed that its viscous evolution may have slowed down due to the inner and outer binaries, and it may be instead evolving mostly through disc winds. Subsequent modeling of the system by \citet{Ronco2021} confirmed that such configuration could indeed result in slower evolution and explain the disc longevity. However, probably the most unique characteristic of HD~98800B is that the disc is in a polar configuration with respect to the inner binary, as revealed by high-resolution ALMA observations at 1.3~mm \citep{Kennedy2019}. Moreover, the A component is expected to transit behind the disc in the immediate future, offering a unique opportunity to characterize the disc structure and composition by monitoring the shape and depth of the transit \citep{Faruqi2025}.

Previous observations with the Karl G. Jansky Very Large Array (VLA) of this system, combined with ancillary data, yielded a low millimetre spectral index between 1~mm and 9~mm for the disc around HD~98800B \citep[$\alpha\approx 2.0-2.1$,][]{Ribas2018}. This low value is a useful diagnostic of grain evolution as it suggests the presence of large, millimetre/centimetre-sized grains, but it can also arise from optically thick emission. Distinguishing between these two possibilities is essential for constraining the dust properties. At 5~cm, the VLA data already revealed significant excess above the extrapolated dust continuum, hinting at additional contributions from free-free and/or gyro-synchrotron emission from disc winds, or stellar magnetic activity \citep{Ribas2018}. The discless A component was also detected at 5~cm, pointing at chromospheric activity. These detections at centimetre wavelengths offer an opportunity to probe stellar activity in the system, and
a robust characterization of their contribution to the observed fluxes is necessary to accurately constrain the dust millimetre spectral index and quantify the level of grain growth in the disc.

In this work, we present new VLA observations of HD~98800 at 6.8~mm (Q~band) and 3~cm (X~band) which complement the previous VLA data of the system at 8.8~mm (Ka~band) and 5~cm (C~band) in \citet{Ribas2018}. Combining these with other ancillary observations, we derive the spectral index of HD~98800 at millimetre/centimetre wavelengths and reveal an azimuthal asymmetry in the system that we interpret as a physical overdensity in the dust distribution. The new VLA observations are first described in Section~\ref{sec:observations}. Section~\ref{sec:results} then presents the imaging process and the obtained fluxes, as well as the modeling approach used to derive the spectral index. Finally, Section~\ref{sec:discussion} discusses the implications for grain growth and possible explanations for the observed asymmetry, and our conclusions are presented in Section~\ref{sec:conclusions}.

\section{Observations}\label{sec:observations}

\subsection{New VLA data}\label{sec:new_observations}

The HD~98800 system was observed by the VLA at wavelengths 6.8~mm (Q~band) and 3~cm (X-band) as part of project 18A-419 (P.I.: Á. Ribas) to complement the previous data at 8.8~mm (Ka~band) and 5~cm (C~band) presented in \citet{Ribas2018}. Both wavelengths were observed in the A configuration (baselines ranging from 0.8 to 36~km) to achieve the highest angular resolution possible.

The Q~band observations were taken on two different dates (15 and 24 April 2018), for a total of 92 minutes on source. The correlator was set up to cover an 8~GHz range between 40 and 48~GHz. The X~band observations consisted of a single execution on 23 March 2018, with a total on source time of 23~minutes. In this case, the correlator set up covers a 4~GHz range between 8 and 12~GHz. In all cases, 3C286 was used as the flux calibrator, J1256-0547 as the bandpass calibrator, and J1127-1857 was the phase calibrator.

The data were calibrated using CASA version 6.6.6~\citep{CASA} and the VLA pipeline \citep{VLApipeline} version 2025.1.0.32. In the case of the Q~band observations, antennas {\tt ea08} and {\tt ea24} suffered from a hardware issue affecting spectral windows 50-65 and we manually flagged them when running the pipeline. For both the Q~band and X~band data sets, we performed a single round of phase-only self-calibration by combining all the spectral windows for the whole duration of the observation. This process increased the peak signal to noise from 9 to 13 at 6.8~mm, and from 42 to 49 for the 3~cm data.

\subsection{Ancillary observations and spectral energy distributions}

HD~98800 has been observed extensively at multiple wavelengths and with various instruments and telescopes. In particular, for this work we also make use of the previous VLA observations at Ka~band (8.8~mm) and C~band (5~cm) analyzed in \citet{Ribas2018}, as well as the ALMA Band 6 (1.3~mm) observations presented in \citet{Kennedy2019}. In addition, the spectral energy distributions (SEDs) of the A and B components of the system are populated from the optical to centimetre wavelengths. Here we make use of the SED data compiled in \citet{Ribas2018}, complemented with the 1.3~mm ALMA flux \citep[47$\pm$5~mJy,][]{Kennedy2019}, as well as a 3~mm flux of HD~98800B from archival data (ALMA project 2016.1.01042.S, P.I.: C. Chandler). We estimated a 3~mm flux of $10.6\pm0.5$~mJy by fitting a Gaussian to the emission, as the source appears unresolved. The uncertainty includes a 5~\% absolute flux calibration error. We note that we update the uncertainty of the SMA flux at 880~$\mu$m in \citet{Andrews2010a} to include a 10\% absolute uncertainty.

\section{Results}\label{sec:results}

\subsection{VLA images, fluxes, and dust mass}\label{sec:results_im_and_fluxes}

\begin{figure*}
	\includegraphics[width=\hsize]{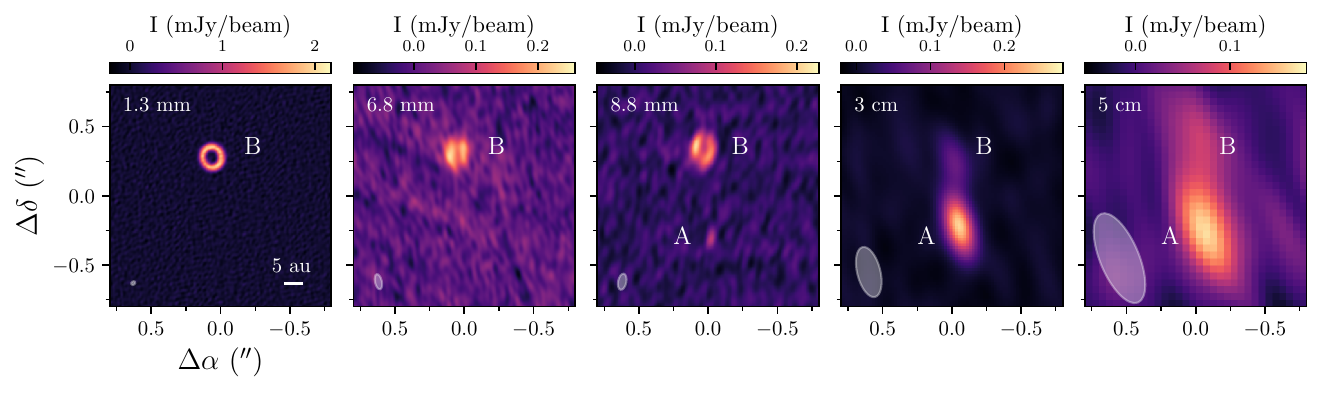}
    \caption{Gallery of millimetre/centimetre observations of HD~98800. The 1.3~mm image is from ALMA \citep{Kennedy2019}, while the rest of them are VLA observations presented here (6.8~mm and 3~cm) and in \citet[][8.8~mm and 5~cm]{Ribas2018}. The images have been aligned to correct for the proper motion of the system. The synthesized beams of the observations are shown on the bottom left corners. In each panel, the A and B components are marked if detected.}
    \label{fig:obs_gallery}
\end{figure*}

The new VLA observations were imaged using the {\tt tclean} task in CASA version 6.6.6. In the case of the Q~band, we used {\tt natural} weighting to maximize sensitivity, while for the X~band we used {\tt briggs} weighting \citep{Briggs1995} with a {\tt robust} parameter of 0.5 to maintain a good sensitivity while also resulting in a better angular resolution. For both wavelengths, the {\tt hobgom} deconvolver was used. The resulting images are shown in Fig.~\ref{fig:obs_gallery}, together with the previous ALMA 1.3~mm and VLA 8.8 and 5~cm observations. The updated SEDs can be found in Appendix~\ref{sec:appendix_newSEDs}. The VLA fluxes for both components \citep[both from this work and from][]{Ribas2018} and the new 3~mm ALMA flux are summarized in Table~\ref{tab:VLA_fluxes}.

\begin{table}
	\begin{center}
	\caption{VLA flux densities of HD~98800A and B from this work and from \citet{Ribas2018}.}
	\label{tab:VLA_fluxes}
	\begin{tabular}{lcl}
		\hline
		Wavelength (mm) & Flux ($\mu$Jy) & Reference\\
		\hline
        \multicolumn{3}{c}{HD~98800A}\\
        \hline
        2.9 (ALMA B3) & $<$ 170$^{\dagger}$  & This work \\
        6.8 (Q band)  & $<$ 60$^{\dagger}$   & This work \\
        8.8 (Ka band) & 80$\pm$15            & \citet{Ribas2018} \\
        30 (X band)   & 251$\pm$15           & This work \\
        50 (C band)   & 160$\pm$30           & \citet{Ribas2018} \\
		\hline
        \multicolumn{3}{c}{HD~98800B}\\
        \hline        
        2.9 (ALMA B3) & 10600$\pm$500  &  This work\\
        6.8 (Q band)  & 1110$\pm$120   &  This work \\
        8.8 (Ka band) & 810$\pm$90     &  \citet{Ribas2018} \\
        30 (X band)   & 73$\pm$10      &  This work \\
        50 (C band)   & 90$\pm$10      &  \citet{Ribas2018} \\
		\hline
	\end{tabular}
    \end{center}

    $^{\dagger}$3$\sigma$ upper limit.
\end{table}

The disc around HD~98800B is clearly detected at 6.8~mm. Although the resulting beam is significantly elongated ($0\farcs11\times0\farcs05$, PA=12~$\deg$), the disc inner cavity and outer radius are resolved in the image, and are fully compatible with values derived from the ALMA high angular resolution data \citep{Kennedy2019}. As in the 8.8~mm image in \citet{Ribas2018}, the eastern side of the disc appears brighter than the western one. We measured a flux of 1110$\pm$120~$\mu$Jy at 6.8~mm, where the uncertainty includes the image root mean square (rms) of 19~$\mu$Jy/beam as well as the 10\% absolute flux calibration error at Q~band. In the case of the A component, there is a very marginal ($\sim$3$\sigma$) possible detection at the expected position of the star, but it only appears after the self-calibration and is comparable to other structures in image noise, so we do not consider it a reliable detection.

Under the assumption of optically thin emission, the dust mass of protoplanetary discs can be estimated from their millimetre fluxes \citep{Hildebrand1983,Beckwith1990} following:
\begin{equation}
    M_{\rm dust} = \frac{F_\nu d^2}{\kappa_\nu B_\nu (T_{\rm dust})}
\end{equation}
where $M_{\rm dust}$ is the total dust mass, $F_\nu$ is the flux density observed at the frequency $\nu$, $d$ is the distance to the source (which we set to 45~pc), $\kappa_\nu$ is the dust opacity, and $B_\nu (T_{\rm dust})$ is the blackbody emission for a dust temperature $T_{\rm dust}$. Estimates using this approach are highly uncertain given the poor constraints on some of the quantities involved, but they are nonetheless useful for comparisons among different discs. To account for these  uncertainties, we bootstrapped 1000 estimates by randomly varying the 6.8 mm VLA flux within its uncertainty following a Gaussian distribution, the dust opacity at 6.8~mm between 0.03 and 0.3~g~cm$^{-2}$ uniformly in logarithmic space \citep[based on estimates for multiple discs with densely populated SEDs presented in][]{Painter2025}, and the dust temperature uniformly between 50 and 100~K. This temperature range is higher than the standard 20~K adopted for more extended discs, as compact discs such as the one around HD~98800B will be hotter, and it is also consistent with the results of the radiative transfer models in Sec.~\ref{sec:opt_thick_disc}. This yielded a dust mass of $2.7^{+3.4}_{-1.4} \times 10^{-5} M_\odot$, which is comparable to other protoplanetary discs \citep{Manara2023}. However, we caution that this mass value was calculated using a higher temperature (most discs are colder), and a different opacity law than most estimates in the literature (dust measurements are usually performed at shorter wavelengths). In addition, some of the 6.8~mm emission could still be optically thick, and thus the derived dust mass may constitute a lower limit to the true value.

In contrast to the 6.8~mm data, the 3~cm image shows a strong detection of the discless HD~98800A component, and a fainter one corresponding to HD~98800B. Due to the lower angular resolution at this wavelength (the beam of the {\tt robust=0.5} image is 0.37$\arcsec \times$0.17$\arcsec$, PA=17 $\deg$), both sources appear unresolved and are slightly blended. We therefore calculated the flux of the A and B components by fitting two point sources with the {\tt imfit} task in CASA. We obtained 3~cm fluxes of 251$\pm$15~$\mu$Jy and 73$\pm$10~$\mu$Jy for HD~98800A and B, respectively, where the uncertainties combine those from {\tt imfit} and a 5\% absolute flux calibration uncertainty at X~band. The rms of the image is 5~$\mu$Jy/beam.

An interesting feature of both the 6.8~mm and 8.8~mm VLA images of HD~98800B is the presence of an azimuthal asymmetry in the disc emission, with the eastern side being brighter than the western one ($\sim$15\% brighter at 6.8~mm, and $\sim$25\% at 8.8~mm). We calculate the statistical significance of the asymmetry as $(I_{\rm max, East} - I_{\rm max, West})/\sqrt{2 \sigma_{\rm rms}^2}$, where $I_{\rm max, East}$ and $I_{\rm max, West}$ are the emission peaks on the east and west side of the disc, and $\sigma_{\rm rms}$ is the corresponding image noise. The resulting significances are 1.0 and 2.8~$\sigma$ at 6.8~mm and 8.8~mm, respectively, which strongly suggests that it is a real feature and not an artifact of the moderate signal to noise of the VLA data (the significance is further increased if we consider that the asymmetry appears on two independent observations). From a direct image inspection, the asymmetry appears to have shifted between both VLA observations from the north-east side of the disc in 2012 to the south-east one in the new 2018 data. This apparent shift is due to the changing orientation of the beam in both data sets (see Section~\ref{sec:opt_thick_disc}). In contrast, the higher resolution 1.3~mm ALMA data from \citet{Kennedy2019} do not show signs of such asymmetry.

\subsection{Spectral indices at millimetre/centimetre wavelengths}\label{sec:spectral_indices}

The new VLA data at 6.8~mm and 3~cm provide additional anchor points to evaluate the spectral index ($\alpha$, where $F_\nu \propto \nu^\alpha$) in the mm/cm wavelength range. We used the measured fluxes of A and B to determine the corresponding spectral indices (calculated simply as $\alpha = \log(F_{\nu_1}/F_{\nu_2}) / \log(\nu_1/\nu_2)$, where $F_{\nu_1}$ and $F_{\nu_2}$ are the fluxes at frequencies $\nu_1$ and $\nu_2$) between various wavelengths using ALMA and VLA data, namely: between 1.3~mm and 6.8~mm,  6.8~mm and the 8.8~mm, 8.8~mm and 3~cm, and 3~cm and 5~cm. The results are summarized in Table~\ref{tab:spectral_indices}.

\begin{table}
    \begin{center}
    \caption{Millimetre and centimetre spectral indices $\alpha$ (where $F_\nu \propto \nu^\alpha$) of HD~98800A and B between different wavelength pairs. }
    \label{tab:spectral_indices}
    \begin{tabular}{lcc}
        \hline
        Component & Wavelengths (mm) & Spectral index $\alpha$ \\
        \hline
        \multirow{2}{*}{HD~98800A}& 8.8 - 30    & $-0.93\pm0.16$ \\
                                   & 30 - 50    & $0.9\pm0.4$ \\
        \hline
        \multirow{4}{*}{HD~98800B} & 1.3 - 6.8  & $2.26\pm0.09$ \\
                                   & 6.8 - 8.8  & $1.2\pm0.6$ \\
                                   & 8.8 - 30   & $1.97\pm0.14$ \\
                                   & 30 - 50    & $-0.4\pm0.3$ \\
        \hline
    \end{tabular}
    \end{center}
\end{table}

To provide a more robust estimate of the spectral index of the B component at various wavelengths and attempt to disentangle the dust emission from other possible mechanisms, we adopt a model with a dust component motivated by \citet{Painter2025}, and a single power law to account for all other emission mechanisms (hereafter referred to as contamination).\footnote{A simpler modeling approach using two power laws (one for the dust contribution and one for the contamination) yielded $\alpha_{\rm dust}=2.09\pm0.05$ when fitting the fluxes from SMA (880~$\mu$m), ALMA (1.3~mm), and the four VLA wavelengths.} The prescription in \citet{Painter2025} (see their equations 13 and 14) uses a function that transitions between two power laws with different spectral indices through a sigmoid function:
\begin{equation}
        F_\nu^{\rm dust} = F_{0,\rm dust} \left(\frac{\nu_{\rm dust}}{\nu_0}\right)^{\eta(\nu_0)} \left(\frac{\nu}{\nu_{\rm dust}}\right)^{\eta(\nu)},
\label{eq:dust} 
\end{equation}
where $F_{0,\rm dust}$ is the flux at a reference frequency \citep[$\nu_0$=33~GHz, following][]{Painter2025}, $\nu_{\rm dust}$ is the frequency at which the emission transitions from optically thick to optically thin, and this transition is regulated by the index $\eta(\nu)$, which is defined as:
\begin{equation}
    \eta(\nu) = \eta_- + (\eta_+ - \eta_-) \left[1 + e^{-\gamma(\nu - \nu_{\rm dust})}\right]^{-1}.
    \label{eq:logistic_transition}
\end{equation}

Here, $\eta_-$ and $\eta_+$ are the spectral indices in the limits $\nu \ll \nu_{\rm dust}$ and $\nu \gg \nu_{\rm dust}$ emission, respectively, which would physically correspond to optically thin and thick dust emission. The parameter $\gamma$ controls the sharpness of the transition between both regimes. The term accounting for the contamination is defined as:
\begin{equation}
F_{\rm cont}(\nu) = F_{0,\rm cont}(\nu/\nu_0)^{\alpha_{\rm cont}}
\label{eq:contamination}
\end{equation}
where $F_{0,\rm cont}$ is the contamination flux at the reference frequency $\nu_0$ and $\alpha_{\rm cont}$ is the corresponding spectral index. The total model is then defined as $F_{\rm total} (\nu)= F_{\rm dust}(\nu) + F_{\rm cont}(\nu)$, and contains seven free parameters: $F_{0,\rm dust}$, $\eta_-$, $\eta_+$, $\nu_{\rm dust}$, $\gamma$, $F_{0,\rm cont}$, and $\alpha_{\rm cont}$. With this model, we fit all existing photometric data at wavelengths $\geq$500~$\mu$m with a Markov Chain Monte Carlo (MCMC) approach using the {\tt emcee} package \citep{emcee}. We used broad priors for all parameters: uniform positive priors for $F_{0,\rm dust}$ and $F_{0,\rm cont}$ explored in logarithmic space (between -4 and -2, and -7 and -4, respectively), uniform priors for $\eta_-$ (between 1.5 and 4) and $\eta_+$ (between 1.5 and 3), a uniform prior for $\log(\nu_{\rm dust})$ between 10 and 1000~GHz (i.e., 0.3~mm to 3~cm), and a broad Gaussian for $\gamma$ centered at -2 with a standard deviation of 0.5 following \citet{Painter2025}. We also enforced $\alpha_{\rm cont} \leq \eta_+ \leq \eta_-$ to prevent swapping between contributions. The results of this fitting procedure are shown in Fig.~\ref{fig:mm_fit} (the corresponding parameter values and corner plot can be found in Appendix~\ref{sec:appendix_modeling}). This process also allows to determine the spectral index of the dust emission as a function of wavelength \citep{Painter2025}:
\begin{equation}
    \alpha_{\rm dust}(\nu) = \eta(\nu) + \frac{d\eta}{d\nu} \nu \log{\left(\frac{\nu}{\nu_{\rm d}}\right)},
    \label{eq:alpha_dust}
\end{equation}

which we show in the bottom panel of Figure~\ref{fig:mm_fit}. The results show preference for a low value of $\alpha_{\rm dust}$ across the millimetre range, with the median below 3 up to $\sim$1~cm. As expected, the spectral index of the contamination component is significantly more negative ($\alpha_{\rm cont}=-0.9^{+0.4}_{-0.5}$). 

\begin{figure}
	\includegraphics[width=\hsize]{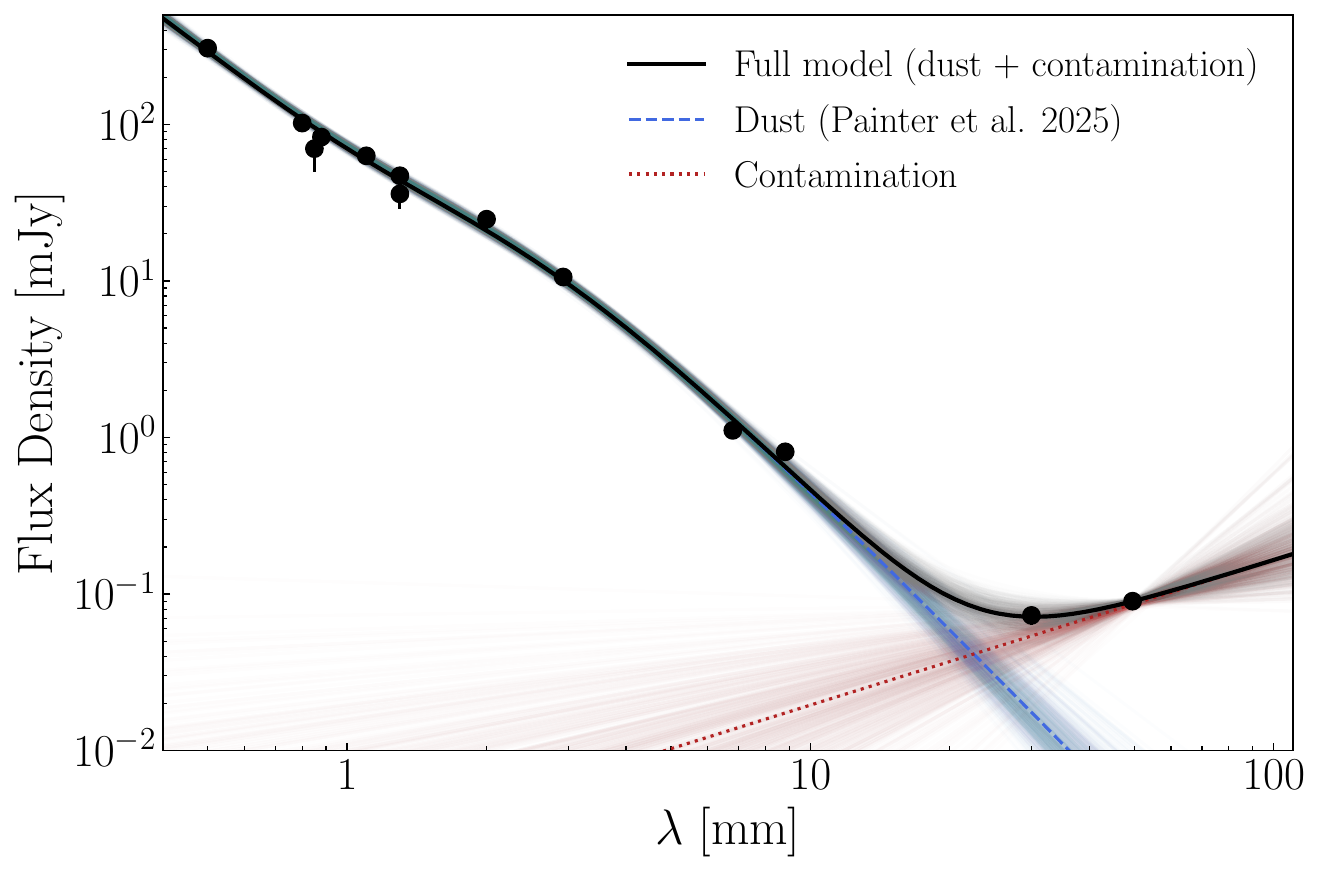}
    \includegraphics[width=\hsize]{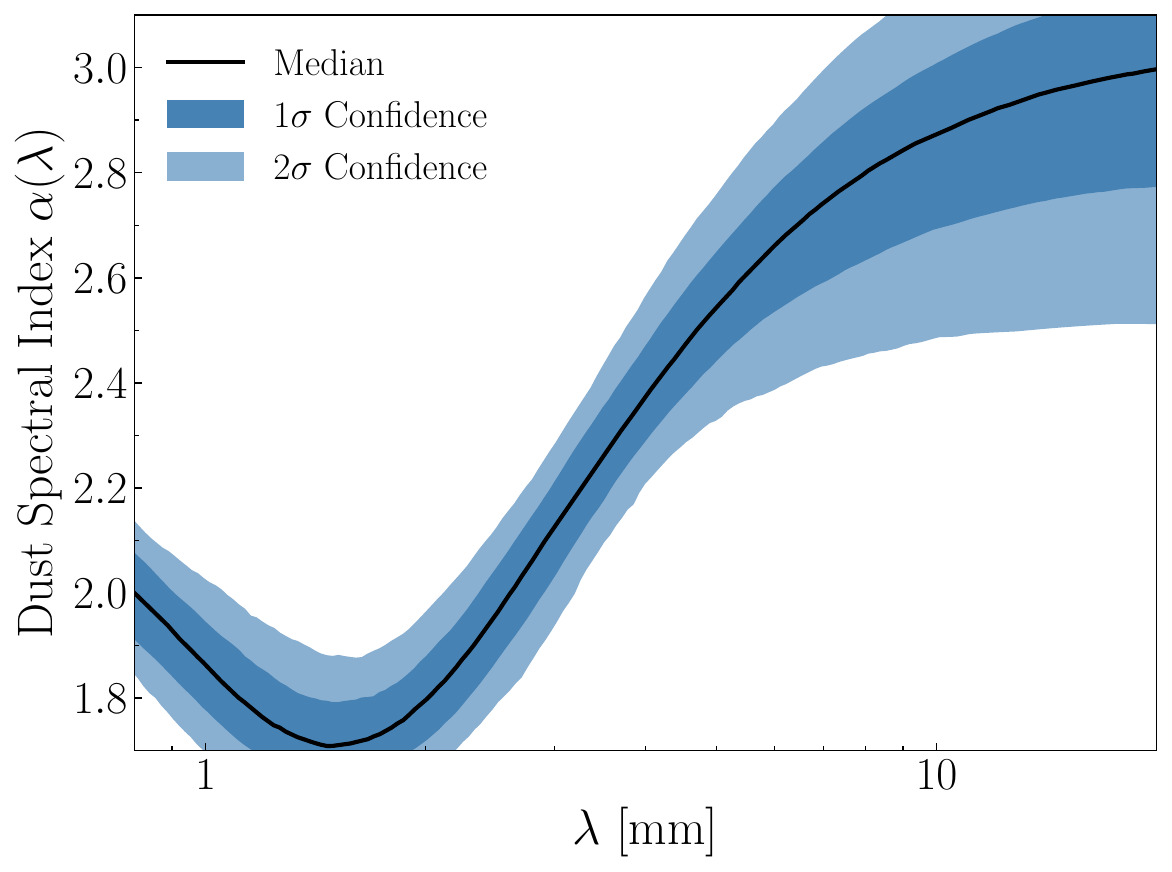}
    \caption{Top panel: Fitting results of the mm/cm SED of HD~98800B including a dust contribution following \citet{Painter2025} and a power law to account for other emission mechanisms. The fit is performed to all the available photometry at wavelengths $\lambda \geq$500~$\mu$m. The solid lines show the median of the total model (black), dust emission (blue), and contamination emission (red), while the shaded regions correspond to 500 models randomly sampled from the posterior distributions. Bottom panel: Dust spectral index $\alpha_{\rm dust}$ as a function of wavelength resulting from the fitting procedure. The black solid line shows the median of the posterior distribution, while the shaded regions correspond to the 1$\sigma$ and 2$\sigma$ confidence intervals.}
    \label{fig:mm_fit}
\end{figure}

\section{Discussion}\label{sec:discussion}

\subsection{Grain growth and non-thermal emission in HD~98800}\label{sec:interpreting_alpha}

The millimetre spectral index is commonly used as a diagnostic of grain growth in protoplanetary discs \citep[e.g.,][]{Ricci2010_Ophiuchus,Tazzari2021_lupus}. For optically thin emission, sub-$\mu$m-sized grains such as those in the interstellar medium and starless cores result in $\alpha=3.5-4$ \citep[e.g.,][]{Draine1984,Schnee2010}, while mm/cm-sized grains have lower $\alpha$ values between $2-3$ \citep[e.g.,][]{Testi2014,Tazzari2021_lupus}. However, the interpretation of low $\alpha$ values in discs is complicated by the fact that their millimetre emission could be (partially) optically thick \citep[e.g.,][]{Beckwith1991,Andrews2005,Zhu2019,Carrasco2019,Ribas2020,Macias2021,Xin2023,Ribas2025}, also resulting in $\alpha$ values close to 2 in the Rayleigh-Jeans regime. Moreover, at sufficiently long wavelengths (typically at $\lambda\approx$1~cm and longer), free-free emission from ionized circumstellar gas \citep[][]{Wright1975,Panagia1975,Reynolds1986} as well as synchrotron emission from the stellar magnetosphere and/or jets \citep[e.g.,][]{Andre1996,Gibb1999,Anglada2018} can also contribute to the total flux density, further complicating the interpretation of $\alpha$ at these long wavelengths.

\citet{Ribas2018} fit the millimetre spectral index of HD~98800B using photometry between 500~$\mu$m and 2.3~mm ($\alpha_{\rm 500~\mu m - 2.3~mm}=2.02^{+0.1}_{-0.05}$) and between 1 and 8.8~mm ($\alpha_{\rm 1~mm - 8.8~mm}=2.09^{+0.14}_{-0.07}$). These $\alpha\sim2$ values are expected for optically thick emission. In contrast, the 5~cm VLA photometry was found to be significantly above the extrapolation of the $1-8.8$~mm spectral index, indicating an additional source of emission at this wavelength other than dust thermal continuum which could potentially affect the fluxes at shorter wavelengths. However, with VLA data at only two different wavelengths, it was not possible to establish how much of the total 8.8~mm emission may be due to non-thermal contamination. \citet{Ribas2018} argued that the low mm spectral index indicated that the HD~98800B is still massive and gas rich, closer in nature to protoplanetary discs instead of debris ones despite its age ($\sim$10~Myr). The ALMA high-resolution observations in \citet{Kennedy2019} clearly confirmed the gas-rich nature of the system, revealing a rotating gaseous disc. Given that grain growth is also to be expected in a $\sim$10~Myr old disc, the cause of the low millimetre $\alpha$ value (optically thick emission, grain growth, or both) in HD~98800B still remained unclear. 

The new VLA spectral indices (Table~\ref{tab:spectral_indices}) confirm the presence of emission mechanisms other than just dust thermal emission at cm wavelengths in the circumstellar disc surrounding HD~98800B. Taken at face value, the steep, negative spectral index between 3 and 5~cm ($\alpha_{\rm 3-5 cm}=-0.4\pm0.3$) is incompatible with dust continuum emission ($\alpha_{\rm dust} \geq2$). It is also unlikely to originate from free-free emission, which typically shows a spectral index $\sim~0.6$ \citep{Panagia1975,Reynolds1986} although it could range between $-0.1$ and 2 in the optically thin and thick extremes, respectively. Instead, this $\alpha_{\rm 3-5 cm}$ value suggests that the emission may be dominated by gyro-synchrotron emission \citep[e.g.,][]{Rodriguez1993,Dzib2013,Ubach2017}, often found in young stellar systems arising from active magnetospheres \citep{Gudel2002}. No jet is known to exist in HD~98800B, as expected for more evolved discs (10~Myr). In addition, \citet{Dzib2013} found that older systems are more likely to display more gyro-synchrotron emission than free-free. We thus interpret the fluxes of HD~98800B at 3 and 5~cm as being due to stellar activity. However, it is also possible that the emission at these wavelengths is, in fact, a combination of free-free and gyro-synchrotron emission. Radio emission from young stars is also known to be highly variable on timescales of hours to days \citep[e.g.,][]{Rivilla2015,Forbrich2017,Ubach2017}, and therefore we cannot unambiguously identify the emission mechanism or estimate the quiescent emission level of HD~98800B with the current data.

Interestingly, HD~98800A also shows clear detections at 8.8~mm, 3~cm, and 5~cm, with fluxes well above the expected photospheric emission. Given that the A component does not host a disc, in this case the origin of this emission is more easily attributed to synchrotron emission from stellar activity. Such emission has been found in various binary systems and their occurence has been linked to their orbital periods in some cases  \citep[e.g.,][]{Massi2006,Salter2010}, possibly due to magnetic reconnection events triggered by the interaction of the stellar magnetospheres.  
Both Aa-Ab and Ba-Bb are close eccentric binaries \citep[$a\sim$1~au, $e=0.5-0.8$,][]{Zuniga2021}, and it is thus possible that their cm emission is induced by a similar process. Given its unique properties, HD~98800 could be a privileged – albeit complex – system to study stellar activity in young stars.

As formerly mentioned, free-free emission from ionized gas in disc winds and/or jets is also common in young stellar systems with circumstellar material \citep[e.g.][]{Macias2016,Anglada2018,Garufi2025}. Free-free emission can also be variable, although on longer timescales than synchrotron emission \citep[e.g.,][]{Ubach2012}. When present, this emission becomes relevant at long mm and cm wavelengths (i.e., it could affect the measured 8.8~mm and even the 6.8~mm flux), and separating it from the dust thermal continuum is crucial to correctly interpret both components \citep[][]{Carrasco2019,Macias2021,Painter2025}. Disentangling these contributions requires a well-populated SED at millimetre and centimetre wavelengths \citep{Painter2025} and, given the scarcity of observations at $\gtrsim$5~mm for the HD~98800 system, fitting three components (dust, free-free, and synchrotron) would be highly degenerate with the existing data. The model in Section~\ref{sec:spectral_indices} is therefore an attempt to quantify the dust emission in the short millimetre range while strongly oversimplifying possible contamination from other emission mechanisms.

With these caveats in mind, the modeling in Section~\ref{sec:spectral_indices} yields low $\alpha_{\rm dust}$ values ($\leq 2.15$) for wavelengths $\lesssim$3~mm. Such low $\alpha_{\rm dust}$ remains indicative of optically thick emission, and also aligns with the general trend that compact discs show, on average, lower spectral indices than their bigger counterparts \citep[e.g.,][]{Tazzari2021_multiwavelength,Chung2024}. We note that the spectral index derived from the modeling in Sec.~\ref{sec:spectral_indices} shows values below 2 up to 3~mm (see Fig.~\ref{fig:mm_fit}), which is typically attributed to optically thick emission from grains with substantial albedo in the millimetre \citep{Zhu2019}. The spectral index increases above 2 at $\lambda \geq 3$~mm and reaches a value $\approx$3 for $\lambda \geq 2$~cm, with uncertainties increasing with wavelength as the contribution from contamination becomes increasingly more important and difficult to disentangle with a sparsely populated SED. We interpret our results as follows:
\begin{enumerate}
    \item The derived dust spectral index indicates that the disc is optically thick up to 2-3~mm, especially given the $\alpha_{\rm dust} < 2$ value.
    \item The increase in the dust spectral index at longer wavelengths imply that the emission is becoming optically thinner.
    \item The fact that the 1.3~mm ALMA data appear mostly azimuthally symmetric but an asymmetry is visible in  the resolved VLA images at 6.8 and 8.8~mm further suggests that the emission from the latter is at least partially optically thin, and the observations are piercing through the disc midplane (see Section~\ref{sec:asymmetry}). It also reinforces the idea that the disc is significantly optically thick at ALMA wavelengths.
    \item If the VLA emission is indeed optically thin, the values of $\alpha_{\rm dust}$ at long millimetre wavelengths indicates the presence of large grains in the disc, corresponding to maximum grain sizes of 0.1-1~mm \citep[e.g.,][]{Testi2014,Birnstiel2018,Guidi2022,Birnstiel2024}.
\end{enumerate}

Despite the additional barriers faced by grain growth in multiple systems \citep[e.g.,][]{Cuello2025}, our modeling suggests that large grains are already present in the disc around HD~98800B and that, despite its age, it still retains a substantial amount of dust. Additionally, it shows that the disc emission remains largely optically thick up to 3 mm, and potentially at even longer wavelengths. Given the unique configuration of HD 98800 – a compact circumbinary disc truncated by external companions — it is unclear whether these high optical depths are a consequence of its specific environment or a common feature of narrow rings. As formerly noted, previous studies have shown that the 1-3~mm emission from discs is at least partially optically thick, particularly in their inner regions. However, this may not apply to rings at larger distances, and many of these seem to have optical depths $\lesssim$1 at these wavelengths \citep[e.g.,][]{Dullemond2018,Macias2021,Zagaria2025}. The disc around HD 98800B could then be the remnant of a larger disc truncated by the A binary, retaining only its inner, optically thick regions. These high optical depths are especially relevant for the study of compact discs that remain unresolved even when observed at high angular resolution \citep[e.g.][]{GuerraAlvarado2025}: if such systems are similarly opaque, then our current dust mass estimates for the compact disc population may be underestimated.

\subsection{The VLA azimuthal asymmetry of the HD~98800B disc}\label{sec:asymmetry}

As formerly mentioned, 6.8~mm and 8.8~mm data show an increase in the disc brightness distribution on its east side. Here we consider three possible explanations for this feature, which we attribute to dust emission. Before we do so, we note that the analysis in the previous section already shows that the emission at 6.8 and 8.8~mm may partially arise from mechanisms other than dust thermal continuum, but we do not expect these to explain the asymmetry. Free-free emission from ionized gas in the disc would appear either as unresolved emission, emission that is azimuthally symmetric distributed in the disc, or perpendicular to it if arising from a jet. In contrast, gyro-synchrotron emission would show as compact emission from the central binary. Therefore, although free-free could contribute to the azimuthal emission at these wavelengths, the asymmetry likely originates from dust continuum emission.

\subsubsection{Illumination changes during the binary orbit}\label{sec:changing_illumination}

The first possible explanation of the asymmetry is that it is due to changes in the illumination of the disc. With a semi-major axis of $a=1$~au and an eccentricity of $e=0.8$ \citep{Zuniga2021}, the polar configuration of the Ba-Bb binary with respect to the disc implies that the stars travel a significant distance above and below the disc midplane during their orbit. This, in turn, results in changes in the stellar flux received by the disc, both temporally and spatially.

This effect was studied in detail by \citet{Rabago2025} by focusing on HD~98800B, finding that the integrated flux emitted by the disc at wavelengths $\gtrsim$20~$\mu$m can vary by up to a factor of two during the orbit depending on the disc vertical structure and its cooling rate. However, the azimuthal brightness distribution of the disc at millimetre wavelengths does not appear to change visibly in their models. We calculated the orbital phase $\phi$ at the time of the VLA 6.8~mm, ALMA~1.3~mm, and VLA 8.8~mm observations using a period of 314.8~days \citep{Zuniga2021}, and using the $T_0$ (passage of periastron) values in \citet{Zuniga2021} ($\phi_{\rm 1.3~mm}=0.81$, $\phi_{\rm 6.8~mm}=0.95$, and $\phi_{\rm 8.8~mm}=0.30$) and \citet{Merle2024} ($\phi_{\rm 1.3~mm}=0.76$, $\phi_{\rm 6.8~mm}=0.9$, and $\phi_{\rm 8.8~mm}=0.24$). These phases reveal that the ALMA 1.3~mm and the VLA 6.8~mm observations were taken at comparable orbital phases and, therefore, the illumination pattern was similar during both observations. Thus, a different illumination pattern cannot account for the the presence of the asymmetry in the 6.8~mm data but the lack of it at 1.3~mm. In addition, both VLA observations were taken before and after the periastron passage but the asymmetry appears on the same side of the disc in both cases, which is also difficult to explain with illumination effects alone. We therefore consider this explanation unlikely.

It is worth noting that the periodic changes in the integrated flux from the disc predicted by \citet{Rabago2025} can have an impact on the study of the HD~98800B disc, as the photometric data used to construct the SED were taken at different epochs and, therefore, different orbital phases. This could affect the spectral indices derived in Section~\ref{sec:results} and introduce additional scatter in the SED that is not accounted for in the photometric uncertainties. However, its 1.3~mm flux appears to be fairly constant within 20\,\% in recent observations, suggesting only a moderate effect in the results.\footnote{While flux measurements at 1.1-1.3~mm display a significant scatter \citep[see][]{Rabago2025}, three ALMA observations at different times taken within a year (two from ALMA project 2016.1.01042.S and one from 2017.1.00350.S) all yielded fluxes between 40-47~mJy.}

\subsubsection{Emission from an optically thick inner wall}\label{sec:opt_thick_disc}

A second possibility is that the asymmetry is due to the emission from the hot inner wall of the disc visible on the far side only. \citet{Ribas2024} showed that many of the crescent-shaped asymmetries observed at millimetre wavelengths in protoplanetary discs appear on their far sides and thus are likely explained by this phenomenon, which also requires the emission to be optically thick.

\begin{figure}
	\includegraphics[width=\hsize]{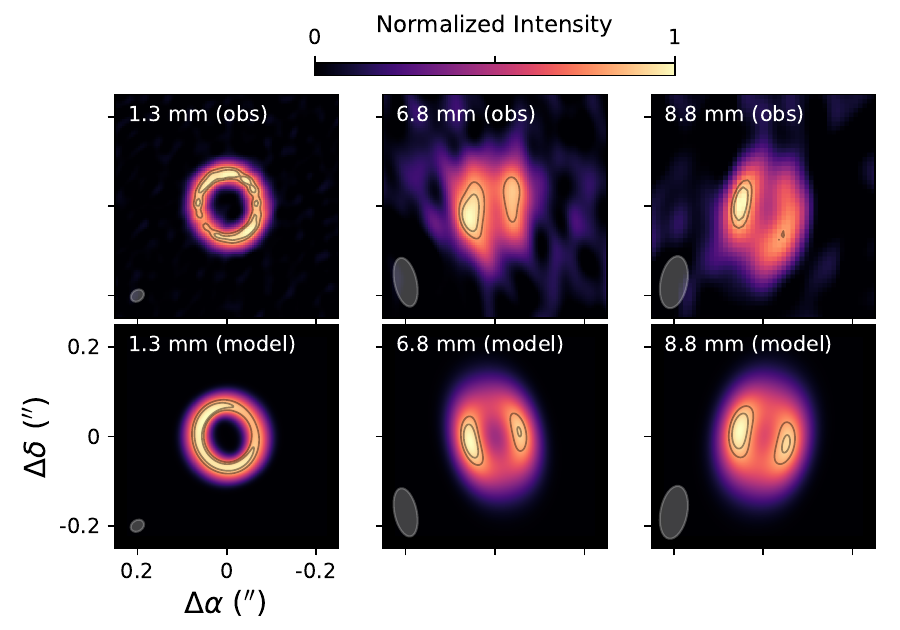}
    \caption{Top row: ALMA and VLA observations of HD~98800B. Bottom row: Simulated MCFOST images (see Section~\ref{sec:opt_thick_disc}) of an optically thick disc with an exposed inner wall, convolved with the corresponding synthesized beams of the observations (shown as gray ellipses). All images are normalized from zero to their peak intensity. Contours correspond to the 80\% and 90\% of the peak intensity. The model successfully reproduces the azimuthal asymmetry observed in the VLA data but also results in an asymmetric disc at 1.3~mm which is brighter on the east side, in contrast with the ALMA observations.}
    \label{fig:simulated_images}
\end{figure}

To show this effect, we produced a radiative transfer model of HD~98800B using the {\tt MCFOST} code \citep{MCFOST,MCFOST2}. We follow the modeling in \citet{Ribas2018}, but update the radial extent (2.5 to 4.6~au), inclination (26~$\deg$), and position angle (16~$\deg$) to match those from the ALMA  high-resolution observations in \citet{Kennedy2019}. We note that the model is quite massive for such a small size ($5\times10^{-5}~M_\odot$ in dust), and its inner region is optically thick even at the VLA wavelengths. We also place the stars 0.5~au above/below the disc plane to better produce a stellar illumination pattern representative of the one experienced by the disc during the binary orbit. In our configuration, the far side of the disc is located on the east side. Even with the stars away from the midplane crossing, the disc still shows an azimuthal asymmetry on the far side. Figure~\ref{fig:simulated_images} show a comparison of the observations at 1.3~mm, 6.8~mm, and 8.8~mm with the simulated images at similar wavelengths and using the corresponding beams. The model qualitatively reproduces the observed asymmetry at 6.8~mm and 8.8~mm.

However, this interpretation has two caveats. First, it requires the east side to be the far side of the disc from Earth for the asymmetry to match the observations. While it was not possible to directly infer the disc orientation from the ALMA images (the existing CO observations only constrain that the north side is rotating towards the Earth), \citet{Kennedy2019} argued that the disc is likely oriented in the opposite sense (with an inclination of 154~$\deg$) based on stability criteria, later also confirmed by the simulations of \citet{Giuppone2019}. The second argument against explaining the VLA azimuthal asymmetry as an inner wall is that the ALMA images do not support this hypothesis: given that the emission at 1.3~mm is optically thicker than at the VLA wavelengths, the asymmetry should be clearly visible at 1.3~mm on the east side too (see Figure~\ref{fig:simulated_images}). This is not the case in the ALMA observations, which instead show two brightness peaks at the disc ansae and tentatively brighter emission on the west side than on the east one. The two peaks at the ansae are likely a result of beam dilution and the ring width being only partially resolved, and our optically thick model also reproduces this effect.\footnote{A similar effect is also present in inclined, optically thin, geometrically thick discs, as the line of sight intercepts more material at the disc ansae \citep{Doi2021}. However, as already discussed, the disc is very likely optically thick at 1.3~mm.} The fact that the west appears slightly brighter in the observations could provide additional support to the idea that it is the far side of the disc, if due to a wall \citep[][]{Ribas2024}. In any case, the ALMA observations rule out the presence of a bright inner wall on the east side, and we therefore discard this explanation as the reason for the asymmetry in the VLA data. 

We do, however, note that the models reproduce the apparent shift in the location of asymmetry. Since dust density structure used in the models is azimuthally symmetric, this is solely due to the different beam orientations in both VLA observations. Therefore, the observations do not imply a physical shift in the azimuthal position of the asymmetry.

\subsubsection{A local overdensity in the dust disc}\label{sec:overdensity}

Another plausible explanation for the azimuthal asymmetry in the VLA observations of HD~98800B is that it corresponds to true overdensity in the dust distribution. In the case of azimuthal asymmetries, vortices \citep{Ataiee2013,Birnstiel2013}, planet/binary-induced horseshoes, \citep[e.g.][]{Ragusa2017,Li2020} and eccentricity gradients \citep{Ragusa2020} may explain the relative abundance of these asymmetries found at millimetre wavelengths \citep{vanderMarel2013,Andrews2018_DSHARP,Ohashi2018,Cazzoletti2018,Riviere-Marichalar2024,Liu2024}.

HD~98800B is a particularly complex system due to the polar configuration of the inner Ba-Bb binary and the influence of the outer A component. The inner binary likely truncates the inner edge of the disc \citep{Ribas2018,Ronco2021}, but it may also induce other structures in the disc such as spiral arms or vortices. Hydrodynamical simulations of circumbinary discs show that the formation of clumps or asymmetries depends sensitively on the binary-disc misalignment: some configurations favour inner dust concentrations \citep{Poblete2019}, whereas others – such as the polar setup explored by \citet{Smallwood2022} – suppress azimuthal asymmetries. In addition, the A component may also influence the disc periodically during its 200-300 year orbit \citep{Zuniga2021}. Simulations of stellar fly-bys in different configurations show that these encounters can induce substructures, which can persist for some time after the flyby has occurred \citep{Cuello2019,Prasad2025}. Similarly, simulations of the HD~98800 system show that passage of the A component around the disc can excite large-scale spiral arms in the disc around B \citep{Smallwood2022,Faruqi2025}. As showcased by these studies, the disc in HD~98800B could be significantly sculped by different mechanisms but, due to its small size, its actual structure is still poorly understood.

The asymmetry in the VLA 6.8~mm and 8.8~mm data has a strong statistical significance, and provides a new piece of information about the disc structure. Since it is hard to explain it through changes in the illumination pattern of the disc or to its inner wall (see Sections~\ref{sec:changing_illumination} and \ref{sec:opt_thick_disc}), we instead attribute it to an actual overdensity in the dust distribution. If this interpretation is correct, then two facts need to be considered and accounted for:
\begin{itemize}
\item The asymmetry is not visible in the ALMA 1.3~mm observations. This could be due to the more efficient trapping of larger pebbles in a local pressure maximum. These large particles emit more efficiently at the VLA wavelengths, and may therefore be only visible in the VLA images. Alternatively, because of their high optical depth, the data at 1.3~mm may not probe deep enough into the disc, hence hiding the asymmetry. Observations at longer wavelengths reach deeper into the disc midplane where large grains could be accumulating, and may thus reveal additional structures \citep[e.g.,][]{Ribas2025}.
\item The asymmetry appears at a similar location in the VLA 6.8~mm and 8.8~mm observations. This implies that either the asymmetry is static, or it was caught by chance at roughly the same location in both observations. Models of gas overdensities induced by eccentricity gradients find that they orbit much slower than the local Keplerian velocity and, therefore, it is possible that the asymmetry has not moved significantly between the 5.5 years between both VLA observations. On the other hand, 5.5~years also happens to be approximately the Keplerian period at 3.5~au for a central source of 1.4~M$_\odot$ \citep[adopting the masses for Ba and Bb in][]{Zuniga2025}. These arguments provide plausible solutions as to why the asymmetry appears on the same location in both VLA data, although we caution that the nature of the asymmetry (i.e., its orbital period and radial distance to the central stars) is unknown.
\end{itemize}

Despite the aforementioned uncertainties, we consider the case for a true overdensity in the disc around HD~98800B to be the strongest among the possible explanations for the azimuthal asymmetry in the VLA data. The complex architecture of the system certainly allows for the presence of localized pressure maxima in the disc such as a vortex, which could account for the observed asymmetry. Given the potential of vortices to enhance planet formation \citep[e.g.,][]{Barge1995,Adams1995,Klahr2006,Eriksson2025} and the configuration and proximity of HD~98800, further observations at millimetre wavelengths are needed to search for possible azimuthal displacements of the asymmetry and better understand its nature. A finer sampling of the millimetre/centimetre SED and observations at even higher resolution, particularly with the ngVLA, will be crucial to fully characterize the properties of this unique system.


\section{Conclusions}\label{sec:conclusions}

This work presents new VLA 6.8 mm and 3 cm continuum observations of the HD~98800 hierarchical quadruple system. Using these and other ancillary data, we have modeled the millimetre/centimetre SED of the disc around HD~98800B and identified an azimuthal asymmetry in it. Our main results are:

\begin{enumerate}
    \item The new VLA data provide a better coverage of the long-wavelength emission of the system. By modeling the emission as a combination of dust thermal continuum and an additional contamination term, we have been able to quantify the dust spectral index to be $\alpha_{\rm dust} \leq 3$ for wavelengths shorter than 1~cm. 
    The spectral index $\alpha<$2 at wavelengths shorter than 3~mm suggests that the emission at those wavelengths is optically thick, while the increasing spectral index at longer wavelengths signals that the emission is becoming optically thinner. The azimuthal asymmetry visible at 6.8~mm and 8.8~mm but not at 1.3~mm provides further evidence for this interpretation. Explaining a spectral index $<3$ up to 1~cm is extremely difficult without the presence of large grains in the disc around HD~98800B.
    \item Both A and B show signatures of emission from mechanisms other than dust thermal emission at centimetre wavelengths. In the case of HD~98800B, the emission at 3-5~cm likely arises from gyro-synchrotron emission, although a contribution from free-free cannot be ruled out. For the discless HD~98800A component, the fluxes at 8.8~mm, 3~cm, and 5~cm can be more robustly attributed to gyro-synchrotron from stellar activity. Since A and B are both close and eccentric binaries ($a \sim1$~au, $e=0.5-0.8$), it is possible that this emission reflects interactions between their magnetospheres.
    \item Both VLA observations of the disc at 6.8 and 8.8~mm show an azimuthal asymmetry, with its east side appearing 15-25\% brighter. In contrast, the ALMA 1.3~mm data do not show signs of this asymmetry. We rule out changes in the illumination pattern during the binary orbit as well as emission from a hot inner wall as plausible explanations for the asymmetry and, instead, interpret it as a real overdensity in the dust distribution, potentially a dust trap induced by a vortex or a relic of the previous passage of the A component. 
\end{enumerate}

Given its unique polar configuration and proximity, HD 98800 stands as a benchmark system for studying planet formation in extreme dynamic environments. The discovery of an azimuthal asymmetry within this disc provides a unique laboratory to test whether vortices or pressure maxima can truly create the conditions for efficient grain growth and planetesimal formation in binary and higher-order multiple systems. Continued characterization of HD 98800B, especially at the resolution offered by the ngVLA, will be essential to fully characterize the complex disc structure, determine the precise nature of the azimuthal asymmetry, and place crucial constraints on the earliest stages of planet formation in this remarkable system.

\section*{Acknowledgements}

We thank the referee for reviewing the manuscript and for their insightful comments and ideas. AR has received funding from the Royal Society through a University Research Fellowship grant number URF\textbackslash R1\textbackslash 241791. AR and CJC have been supported by the UK Science and Technology research Council (STFC) via the consolidated grant ST/W000997/1. A.B acknowledges support from the Deutsche Forschungsgemeinschaft (DFG, German Research Foundation) under Germany's Excellence Strategy – EXC 2094 – 390783311. This project has received funding from the European Research Council (ERC) under the European Union Horizon Europe research and innovation program (grant agreement No. 101042275, project Stellar-MADE). The National Radio Astronomy Observatory and Green Bank Observatory are facilities of the U.S. National Science Foundation operated under cooperative agreement by Associated Universities, Inc. This paper makes use of the following ALMA data: ADS/JAO.ALMA\#2016.1.01042.S, ADS/JAO.ALMA\#2017.1.00350.S. ALMA is a partnership of ESO (representing its member states), NSF (USA) and NINS (Japan), together with NRC (Canada), NSTC and ASIAA (Taiwan), and KASI (Republic of Korea), in cooperation with the Republic of Chile. The Joint ALMA Observatory is operated by ESO, AUI/NRAO and NAOJ.

\section*{Data Availability}

All the observations used in this work are publicly available through the VLA and ALMA archives.




\appendix

\section{Updated spectral energy distributions}\label{sec:appendix_newSEDs}

\begin{figure}
	\includegraphics[width=\hsize]{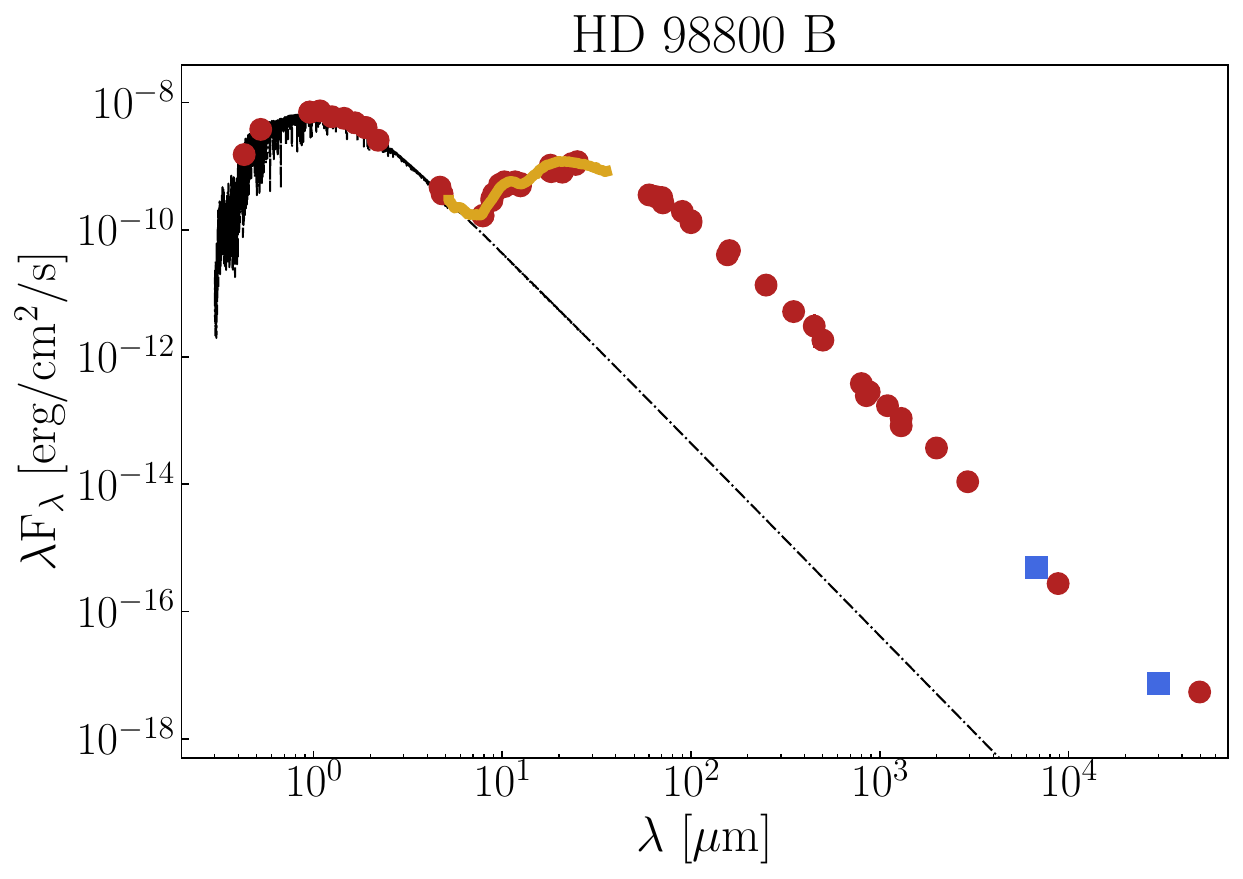}
    \includegraphics[width=\hsize]{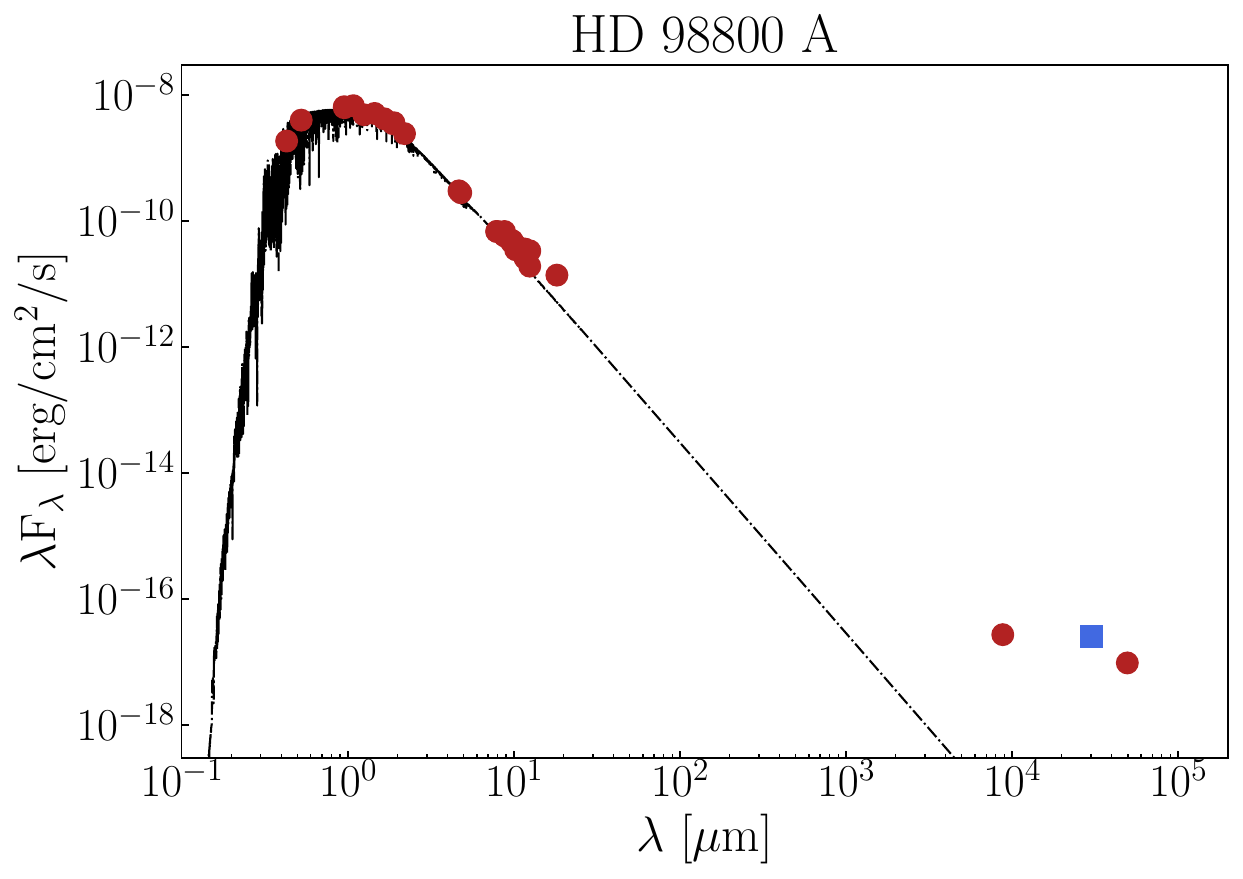}
    \caption{Updated SEDs of HD~98800B (top) and HD~98800A. The new VLA observations are shown as blue squares. The rest of the photometric data compiled in \citet{Ribas2018} are shown as red dots, and the mid-IR spectrum from \emph{Spitzer}/IRS \citep{Furlan2007} is presented in orange. The dashed black lines show the stellar photospheres of a 4200~K and 4500~K star, respectively \citep[following][]{Andrews2010a}.}
    \label{fig:new_SEDs}
\end{figure}

Figure~\ref{fig:new_SEDs} shows the SEDs of both HD~98800A and B, including the VLA flux densities measured in this work.

\section{Modeling results}\label{sec:appendix_modeling}

\begin{figure*}
    \includegraphics[width=\hsize]{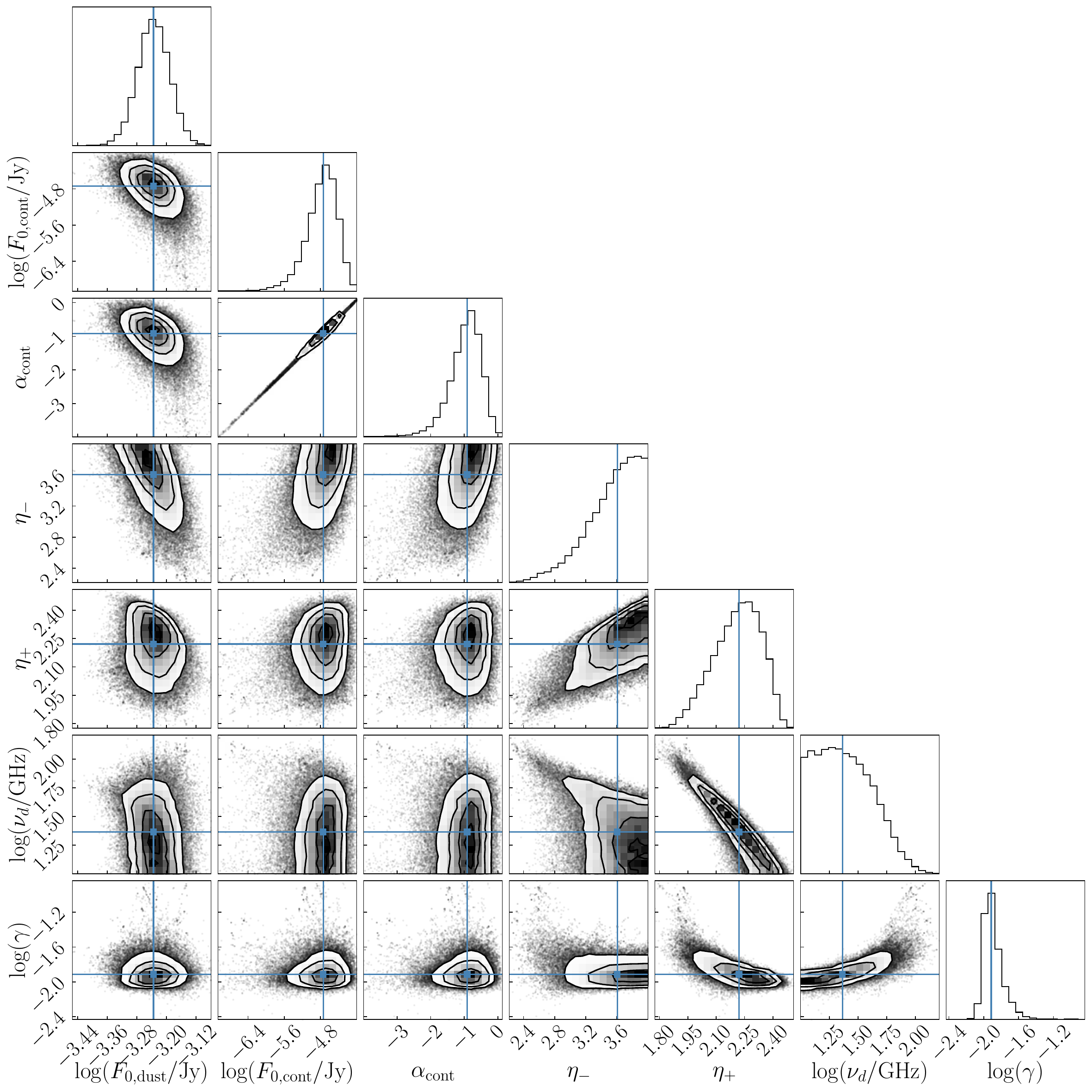}
    \caption{Corner plot showing the posterior distributions of the parameters resulting from the mm/cm SED fitting using two power laws (see Sec.~\ref{sec:spectral_indices} for details).}
    \label{fig:corner}
\end{figure*}

Here we report the parameter values and corner plots of the mm/cm SED fitting  described in Sec.~\ref{sec:spectral_indices}. Table~\ref{tab:params} show the obtained parameters, and Figure~\ref{fig:corner} show the corner plot.

\begin{table}
	\centering
	\caption{Modeling results of the mm/cm SED of HD~98800B as described in Section~\ref{sec:spectral_indices}. The reported values are the median of the posterior distributions, and the uncertainties correspond to the 16th and 84th percentiles.}
	\label{tab:params}
	\begin{tabular}{lc} 
		\hline
		Parameter & Value\\
		\hline
        $\log(F_{0,\rm dust}$/Jy)  & -3.24$^{+0.04}_{-0.04}$ \\
        $\eta_+$                     & 2.22$^{+0.12}_{-0.15}$ \\
        $\eta_-$                     & 3.6$^{+0.3}_{-0.4}$ \\
        $\log(\nu_{\rm dust}$/GHz) & 1.4$^{+0.3}_{-0.3}$ \\
        $\log(\gamma$)             & -1.91$^{+0.13}_{-0.07}$ \\
        $\log(F_{0,\rm cont}$/Jy)  & -4.7$^{+0.3}_{-0.4}$ \\
        $\alpha_{\rm cont}$          & -0.9$^{+0.4}_{-0.5}$ \\
		\hline
	\end{tabular}
\end{table}

\bsp	
\label{lastpage}
\end{document}